\documentclass[usenatbib]{mn2e}
\usepackage{amsmath}
\usepackage{natbib}
\usepackage{epsfig}
\usepackage{txfonts}
\usepackage{color}

\newcommand{\Mpc}{{\rm Mpc}}

\newcommand{\expf}[1]{{{\rm e}^{#1}}}

\newcommand{\zh}{{z_{\rm h}}}

\newcommand{\nbb}{{n^{\rm pl}}}

\newcommand{\nS}{n_{\rm S}}

\newcommand{\id}{{\,\rm d}}

\newcommand{\beq}{\begin{equation}}   %

\newcommand{\eeq}{\end{equation}}   %

\newcommand{\beqa}{\begin{eqnarray}}   %

\newcommand{\eeqa}{\end{eqnarray}}   %

\newcommand{\beal}{\begin{align}}

\newcommand{\enal}{\end{align}}

\newcommand{\bspl}{\begin{split}}

\newcommand{\espl}{\end{split}}

\newcommand{\bsub}{\begin{subequations}}

\newcommand{\esub}{\end{subequations}}

\newcommand{\bmulti}{\begin{multline}}   %

\newcommand{\beqm}{\begin{mathletters}}   %

\newcommand{\eeqm}{\end{mathletters}}   %

\newcommand{\pd}{\partial}

\newcommand{\pAb}[2]{\frac{\displaystyle\pd #1}{\displaystyle\pd #2}}

\newcommand{\Abl}[2]{\frac{{\rm d} #1}{{\rm d} #2}}

\newcommand{\pot}[2]{#1 \times 10^{#2}}

\usepackage{hyperref}
\usepackage{grffile}
\usepackage{graphics}
\usepackage{booktabs}

\title[Green's function]
{Green's function of the cosmological thermalization problem}

\author[Chluba]{J.~Chluba$^{1}$\thanks{E-mail:jchluba@pha.jhu.edu} 
\\
$^{1}$ Johns Hopkins University, Department of Physics and Astronomy, Bloomberg Center 435, 
3400 N. Charles St., Baltimore, MD 21218
}

\begin{document}

\date{{Accepted 2013 June 6. Received 2013 April 10}}

\maketitle

\begin{abstract}
Energy release in the early Universe leads to spectral distortions of the cosmic microwave background (CMB) which in the future might allow probing different physical processes in the pre-recombination ($z\gtrsim 10^3$) epoch. Depending on the energy injection history, the associated distortion partially thermalizes due to the combined action of Compton scattering, double Compton scattering and Bremsstrahlung emission, a problem that in general is hard to solve. Various analytic approximations describing the resulting distortion exist, however, for small distortions and fixed background cosmology the Green's function of the problem can be pre-computed numerically. Here we show that this approach gives very accurate results for a wide range of thermal histories, allowing fast and quasi-exact computation of the spectral distortion given the energy release rate. Our method is thus useful for forecasts of possible constraints on early-universe physics obtained from future measurements of the CMB spectrum.
\end{abstract}

\begin{keywords}
Cosmology: cosmic microwave background -- theory -- observations
\end{keywords}

\section{Introduction}
\label{sec:Intro}
The observations of the cosmic microwave background (CMB) with COBE/FIRAS \citep{Mather1994, Fixsen1996} showed that the CMB frequency spectrum is extremely close to a perfect blackbody, allowing deviations of no more than $\Delta I_\nu/I_\nu \lesssim \pot{\rm few}{-5}$ at wavelengths ranging from $100\,\mu{\rm m}$ to $1\,{\rm cm}$. Dramatic improvements in experimental design and technology over the past decades since the launch of COBE may soon allow much more sensitive measurements of the CMB frequency spectrum \citep[e.g., see][]{Fixsen2002, Kogut2011PIXIE}. 
For this reason, CMB spectral distortion have recently experienced renewed theoretical interest \citep[e.g., see][to name a few]{Chluba2011therm, Chluba2012, Pajer2012, Khatri2012b, Tashiro2012, Yacine2013RecSpec, Sunyaev2013}, potentially providing a way to probe new physics.
It therefore is time to ask what exactly one might hope to learn about the thermal history of our Universe from future measurements of the CMB spectrum and how the information can be extracted.

To address this problem, a fast and accurate method for computing the spectral distortion from different scenarios is required. Previous numerical approaches \citep[e.g., see][]{Procopio2009, Chluba2011therm} are rather time-consuming. Simple analytic approximations exist \citep[e.g., see][]{Sunyaev1970mu, Illarionov1975, Illarionov1975b, Burigana1991, Hu1993, Burigana1995, Khatri2012b, Khatri2012mix}, however, for small distortions and fixed background cosmology it is possible to obtain a numerical approximation for the Green's function of the thermalization problem.
This allows us to compute the spectral distortion for a wide range of thermal histories in an economic and quasi-exact way. In this short paper, we present our method which provides the basis for detailed forecasts, assessing the future experimental possibilities for spectral distortion measurements, and comparisons with independent \citep{Khatri2013forecast} approaches.

\section{Cosmological thermalization problem}
\label{sec:thermalization_problem}
In the early Universe, photons undergo many interactions with the electrons and baryons. A detailed discussion of the cosmological thermalization problem for different energy-release mechanisms can be found in \citet{Chluba2011therm}. 
Compton scattering allows photons to exchange energy with the electrons and baryons, causing diffusion in frequency, while double Compton and Bremsstrahlung emission are responsible for adjusting the photon number.
The Boltzmann equation for the evolution of the average photon occupation number, $n_\nu(t)$, can be expressed as:
\beq\label{eq:BoltzEq_Photons}
\pAb{n_{\nu}}{t}-H\,\nu\,\pAb{n_{\nu}}{\nu}
=\left.\Abl{n_{\nu}}{t}\right|_{\rm C}
+\left.\Abl{n_{\nu}}{t}\right|_{\rm DC}
+\left.\Abl{n_{\nu}}{t}\right|_{\rm BR}
+\left.\Abl{n_{\nu}}{t}\right|_{\rm S}.
\eeq
The second term on the left hand side describes the redshifting of photons due to the adiabatic expansion of the Universe and the right hand side terms correspond to the physical processes quoted above. 
We also explicitly added a photon source term, $\left.\id n_{\nu}/\id t\right|_{\rm S}$, however, direct photon production is normally negligible. If, for example, energy release is caused by some decaying or annihilating relic particle, very few soft photons are directly produced, and high-energy photons and decay products just heat the medium \citep[e.g., see][]{Shull1985, Chen2004, Huetsi2009, Slatyer2009, Cirelli2009}. 
This leads to up-scattering of CMB photons and thus sources distortions, initiating the cosmological thermalization process.

In addition to the photon occupation number, one has to follow the temperature of ordinary matter, $T_{\rm m}$, including the cooling/heating due to the adiabatic expansion of the Universe, Compton scattering, double Compton emission and Bremsstrahlung \citep[for details see][]{Chluba2011therm}:
\beq\label{eq:Te_eq}
\frac{\id T_{\rm m}}{\id t}
=- 2 H T_{\rm m} 
+\left.\Abl{T_{\rm m}}{t}\right|_{\rm C}
+\left.\Abl{T_{\rm m}}{t}\right|_{\rm DC/BR}
+\frac{\dot{Q}(t)}{k \alpha_{\rm h}},
\eeq
where $k \alpha_{\rm h}= (3k/2)[N_{\rm e} + N_{\rm H} + N_{\rm He}]$ denotes the heat capacity of an ideal electron, hydrogen and helium plasma.
Extra energy release can be parametrized using an effective heating rate, $\dot{Q}(t)$. It increases the matter temperature, subsequently causing up-scattering of CMB photons by Compton scattering and photon production at very low frequencies \citep{Sunyaev1970mu}.

For us it is important that any primordial spectral distortion has to be {\it tiny} \citep[cf.,][]{Fixsen1996, Fixsen2009}, so that the thermalization problem can be linearized\footnote{We numerically solve the full non-linear system, however, limiting the total energy release to $\Delta\rho_\gamma/\rho_\gamma \ll 1$.}. We furthermore know the main cosmological parameters, such as the baryon density or Hubble expansion rate, extremely well from measurements of the CMB anisotropies \citep{WMAP_params, Planck2013params}. Thus it is possible to numerically compute the Green's function, $G_{\rm th}(\nu, z_{\rm h}, z_{\rm obs})$, of the thermalization problem using {\sc CosmoTherm}. Here $z_{\rm h}$ denotes the heating redshift and $z_{\rm obs}$ the observing redshift. The resulting spectral distortion at $z=0$ is then given by 
\beq\label{eq:final_dist}
\Delta I_\nu(z=0)
=\int G_{\rm th}(\nu, z', 0) \, \frac{\id (Q/\rho_\gamma)}{\id z'} \id z',
\eeq
where $\rho_\gamma(z)$ is the energy density of the CMB photon field, and we introduced the intensity, $\Delta I_\nu\equiv (2h\nu^3/c^2)\Delta n_\nu$. We chose to use redshift as time variable, since most energy release mechanisms have rather simple form in this coordinate \citep[see][]{Chluba2011therm}. From the practical point of view this poses no significant limitation, and Eq.~\eqref{eq:final_dist} can be generalized to other variables, as long as the characteristic time over which the heating rate varies is longer than the width of the narrow Gaussian used to approximate the $\delta$-function energy release (see next section).

\begin{figure*}
\centering
\includegraphics[width=1.5\columnwidth]{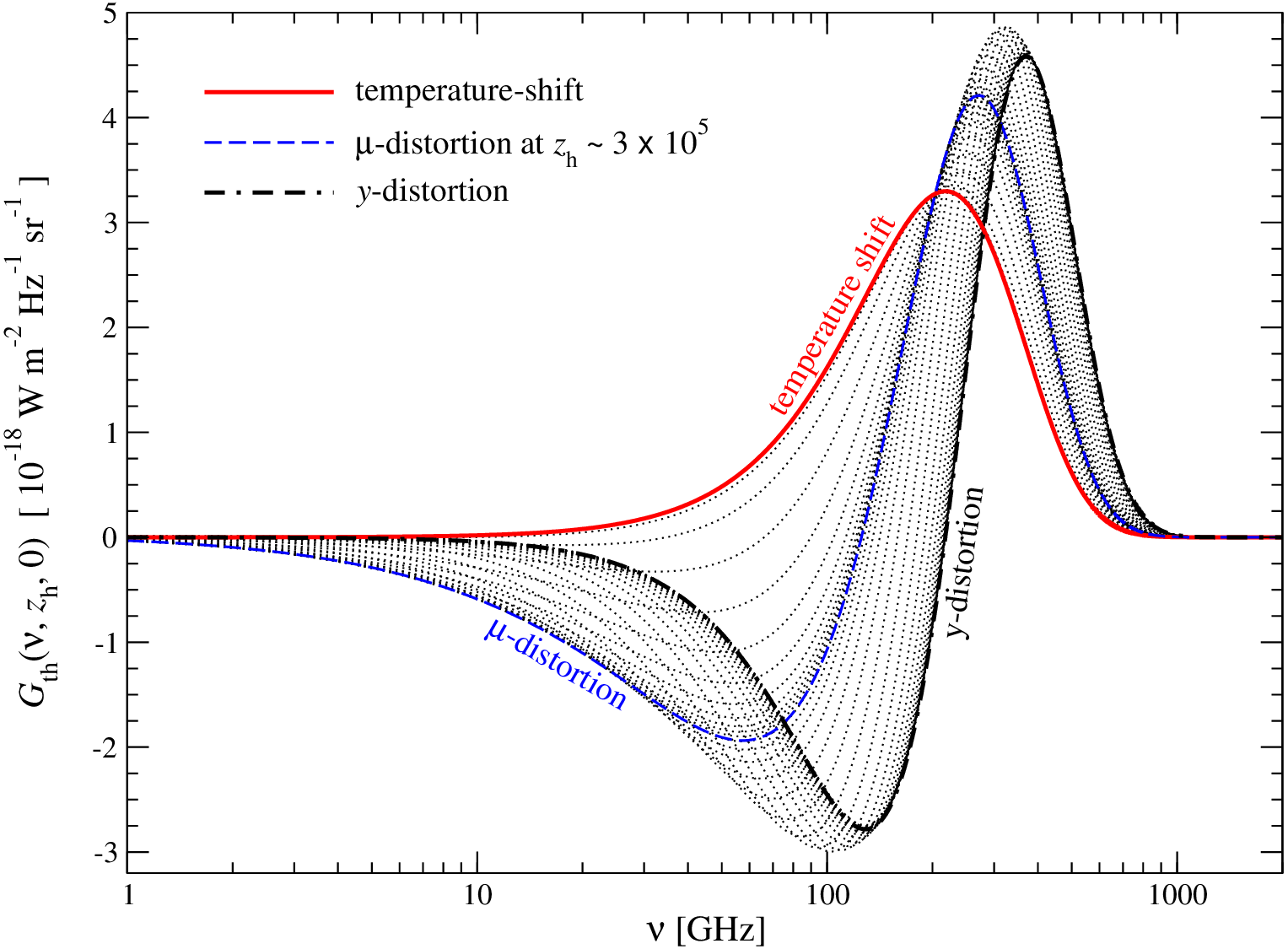}
\caption{Numerical results for the Green's function of the cosmological thermalization problem for various heating redshifts, $z_{\rm h}\in [10^3, \pot{5}{6}]$. Energy release at very high redshifts causes an increase in the effective temperature of the photon field, while at low redshifts photons only partially up-scatter, creating a $y$-distortion. Around $z_{\rm h}\simeq \pot{3}{5}$ a pure $\mu$-distortion is created. All intermediate stages are roughly (precision $\simeq 10\%-30\%$) represented by a superposition of these extreme cases, however, the small residuals provide a principle possibility to distinguish different thermal histories at redshifts $10^4\lesssim z\lesssim \pot{3}{5}$.}
\label{fig:Greens}
\end{figure*}

\section{Results for the Green's function}
\label{sec:Greens}
To obtain the Green's function we solve the thermalization problem assuming that the heating rate is given by a $\delta$-function centered at heating redshift, $\zh$. Numerically we simply approximate the heating rate by a narrow Gaussian in redshift
\beq\label{eq:delta_appr}
\left.\frac{\id (Q/\rho_\gamma)}{\id z}\right|_{\delta} \approx \frac{\Delta \rho_{\gamma, \rm h}}{ \rho_{\gamma, \rm h}}\,\frac{\exp\left(-[z-z_{\rm h}]^2/2\sigma_z^2\right)}{\sqrt{2\pi\sigma_z^2}}
\eeq
with\footnote{We confirmed the convergence of the Green's function by changing the width of the Gaussian, finding excellent agreement between the results.} $\sigma_z\simeq 0.01$ and $\Delta \rho_{\gamma, \rm h}/\rho_{\gamma, \rm h}\simeq 10^{-6}$. By evolving the distortion, $\Delta n_\nu$, and matter temperature, $T_{\rm m}$, from starting redshift $z_{\rm s}\gg z_{\rm h}$ until late times ($z\simeq 200$) we obtain an approximation for the Green's function from $G_{\rm th}(\nu, z_{\rm h}, 0)\approx \Delta n^{\rm num}_\nu/[\Delta \rho_{\gamma, \rm h}/\rho_{\gamma, \rm h}]$. At $z_{\rm s}$ the photon field is given by a blackbody at temperature $T_{\rm s}<T_0(1+z_{\rm s})$, where $T_0=2.726\,{\rm K}$ \citep{Fixsen2009} is the CMB temperature today.
Initially, also the temperature of ordinary matter is $T_{\rm m}(z_{\rm s})\equiv T_{\rm s}$. At the end of the evolution, we ensure that the effective photon temperature is close to $T_0$. To compute the Green's function we furthermore exclude the down-scattering of photons by adiabatically cooling electrons, leading to a distortion characterized by {\it negative} chemical potential and $y$-parameter \citep{Chluba2005, Chluba2011therm, Khatri2011BE}. This process can be included afterward, since for fixed cosmology the corresponding distortion is known precisely.
We adopted the following cosmological parameters: $Y_{\rm p}=0.24$, $\Omega_{\rm m}=0.26$, $\Omega_{\rm b}=0.044$, $\Omega_{\Lambda}=0.74$, $\Omega_{\rm k}=0$, $h=0.71$, and $N_{\rm eff}=3.046$, however, since the computation is rather straightforward, it is very easy to adjust these values.

\subsection{Numerical results}
In Fig.~\ref{fig:Greens} we present the numerical results for the Green's function of the cosmological thermalization problem. 
For very early energy release ($\zh\gg z_\mu\simeq \pot{2}{6}$), it is equivalent to a temperature shift, having the shape
\beq\label{eq:Temp}
\nonumber
4 G_{\rm th}(\nu, z_{\rm h}\gg z_\mu, 0)\approx G(\nu)=\frac{2h\nu^3}{c^2}\,\nu \partial_\nu \nbb(\nu, T_0)=\frac{2h\nu^3}{c^2}\frac{x\,\expf{x}}{(\expf{x}-1)^2},
\eeq
where $x=h\nu/kT_0$, $\nbb(\nu, T_0)=1/(\expf{x}-1)$, and the factor of $4$ arises from $\Delta \rho_\gamma/\rho_\gamma\simeq 4 \Delta T/T $. This indicates that the thermalization process is extremely efficient, with the residual $\mu$-type spectral distortion \citep{Sunyaev1970mu} being exponentially suppressed by the distortion visibility function\footnote{See \citet{Khatri2012b} for recent analytical discussion.}, $\mathcal{J}(\zh)=\expf{-(\zh/z_\mu)^{5/2}}$ \citep{Danese1982, Hu1993}. 
To improve the accuracy of our computation, at $z\gtrsim \pot{5}{4}$ we treated the temperature shift independently, imposing photon number conservation, $\int \nu^2 \Delta n_\nu \id \nu \equiv 0$, to define the distortion part\footnote{The thermalization process slowly increases the photon number density or equivalently the effective temperature (defined by $N_\gamma\propto T^3$). To avoid having a {\it large} contribution from $G(\nu)$ to the numerical solution we can restart the computation at different stages, shifting the temperature of the reference blackbody to the new effective temperature of the photon field.  For practical purposes we require that the overall contribution of $G(\nu)$ to the solution of $\Delta I_\nu$ remains less than $\simeq 0.01\%$. At $z\lesssim\pot{5}{4}$ thermalization is already very inefficient and no significant {\it average} temperature shift can be produced, so that at that point we leave the reference blackbody unchanged.}. Since the temperature shift caused by early energy release is not directly observable, this procedure allowed us to obtain precise results for the $\mu$-type distortion related to significant energy release at very early times. 
Large energy release, $\Delta \rho_\gamma/\rho_\gamma\simeq 10^{-2}$ say, at $z\gg \pot{2}{6}$ is in principle still possible without violating {\it COBE}/{\it FIRAS} spectral distortion limits since the final amplitude of the $\mu$-distortion is exponentially suppressed by the distortion visibility function. Given the limiting chemical potential, $\mu_{\rm lim}\ll 1$, to which the experiment might be sensitive, and allowing at most $q_{\rm max}=\left.\Delta \rho_\gamma/\rho_\gamma\right|_{\rm max}$ energy-release at a single redshift one can in principle probe until redshift $z_{\rm lim}\simeq \pot{2}{6}\ln\left(1.4 q_{\rm max}/\mu_{\rm lim}\right)^{2/5}$. For $q_{\rm max}=10^{-2}$ and $\mu_{\rm lim}=10^{-8}$ this means $z_{\rm lim}\simeq \pot{5.8}{6}$. From the numerical point of view, our calculation should still be valid to $\lesssim 0.1\%$ precision in this case, although the final distortion is suppressed by about six orders of magnitude relative to the initial value.

For low heating redshift ($\zh\lesssim 10^4$), the Green's function takes the shape of a Compton $y$-distortion \citep{Zeldovich1969}. At this redshift, the effective $y$-parameter remains very small, so that photons only weakly interact with the electrons and the distortion is described by
\beq\label{eq:YSZ}
\nonumber
4 G_{\rm th}(\nu, z_{\rm h}\lesssim 10^4, 0)\approx Y_{\rm SZ}(\nu)=\frac{2h\nu^3}{c^2}\frac{x\,\expf{x}}{(\expf{x}-1)^2}\left[ x\coth(x/2)-4\right].
\eeq
At very low frequencies ($\nu\lesssim 1\,{\rm GHz}$) this approximation is not as accurate, since matter and radiation start decoupling and Bremsstrahlung emission (or absorption) can alter the spectrum \citep[see][for illustration]{Chluba2011therm}. However, the photon intensity in that part of the radiation field is very small and this effect will be challenging to observe. The numerical results for the Green's function include these aspects.

At $z_{\rm h}\simeq \pot{3}{5}$ the Green's function is mainly represented by a $\mu$-type distortion, obtained using the condition $\int \nu^2 \Delta n_\nu \id \nu \equiv 0$ \citep[see][for more detail]{Chluba2012, Khatri2012mix}:
\beq\label{eq:mu_dist}
\nonumber
\alpha^{-1} G_{\rm th}(\nu, z_{\rm h}\simeq \pot{3}{5}, 0)\approx M(\nu)=\frac{2h\nu^3}{c^2}\frac{\expf{x}}{(\expf{x}-1)^2}\left[ x/\beta -1\right].
\eeq
Here $\alpha \approx 1.401$ \citep{Sunyaev1970mu, Illarionov1975, Illarionov1975b} and $\beta=3\zeta(3)/\zeta(2)\approx 2.1923$. Neglecting the temperature shift, the Green's function for the distortion part is well represented by $M(\nu)$, even for $\zh\gtrsim \pot{3}{5}$ when additionally multiplying with the distortion visibility function, $\mathcal{J}(\zh)$, giving the approximation $G_{\rm th}(\nu, z_{\rm h}\gtrsim \pot{3}{5}, 0)\approx 1.401\mathcal{J}(\zh)\,M(\nu)+\frac{1-\mathcal{J}(\zh)}{4}\,G(\nu)$.

\begin{figure*}
\centering
\includegraphics[width=1.4\columnwidth]{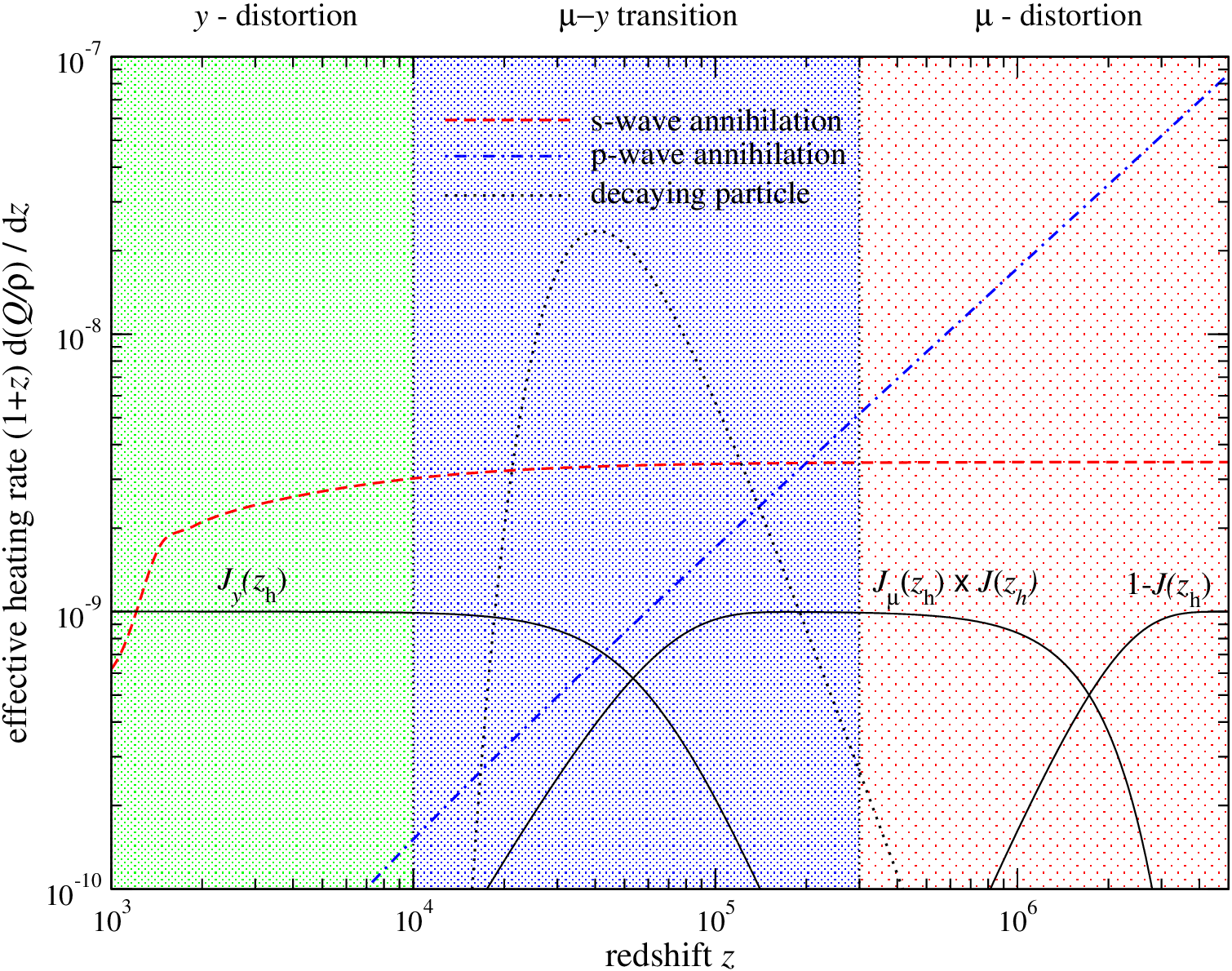}
\\[2mm]
\includegraphics[width=1.4\columnwidth]{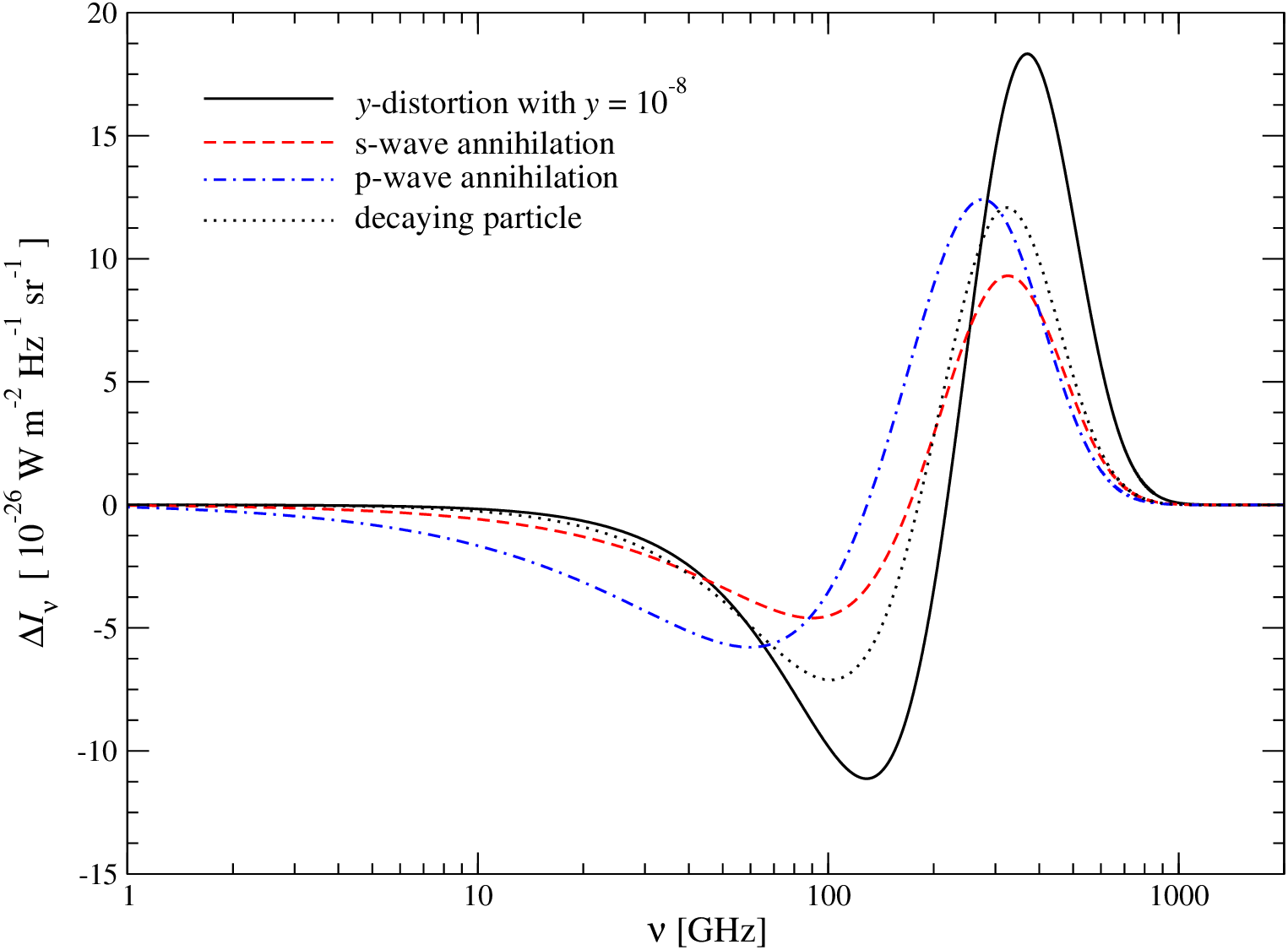}
\caption{Effective heating rate and spectral distortion created by different scenarios. In all cases the total energy release was similar to $\Delta\rho_\gamma/\rho_\gamma\simeq {\rm few}\times 10^{-8}$ (irrespective of other observational constraints). We only included the distortion part for the calculation (i.e., set the $G(\nu)$-term to zero) since the shift in the overall temperature of the photon field remains unobservable. For comparison we show a $y$-distortion with $y\simeq 10^{-8}$. We also illustrate the redshift-dependence of $\mathcal{J}_{y}(\zh)$, $\mathcal{J}_{\mu}(\zh)$ and $\mathcal{J}(\zh)$, which we multiplied by $10^{-9}$, for convenience. The different scenarios are described in more detail in Sect.~\ref{sec:examples_diss}.}
\label{fig:examples}
\end{figure*}

Finally, Figure~\ref{fig:Greens} also shows the Green's function at several intermediate stages, $10^4 \lesssim \zh\lesssim \pot{3}{5}$. Signals produced mainly during this epoch were already discussed in \citet[][see Fig.~15 and 19]{Chluba2011therm}, indicating that the total distortion is not simply given by a pure superposition of $\mu$ and $y$-distortion. More recently, this was also demonstrated by \citet{Khatri2012mix}.
The small ($\simeq 10\%-30\%$) residuals might allow distinguishing different energy release scenarios in the future. This is especially interesting if, for instance, decaying particles with lifetimes $t_{\rm X} \simeq \pot{2.6}{8}\,{\rm sec}-\pot{2.2}{11}\,{\rm sec}$ are present \citep{Chluba2011therm}, however, detailed forecasts are required to address this question.
One can still obtain a fairly good approximation for the Green's function in this intermediate regime by assuming that the spectrum is close to a superposition of a pure $\mu$- and $y$-distortion, $G_{\rm th}(\nu, z_{\rm h}\lesssim\pot{3}{5}, 0)\approx 1.401\mathcal{J}_{\mu}(\zh)\,M(\nu)+\frac{\mathcal{J}_{y}(\zh)}{4}\,Y_{\rm SZ}(\nu)$.
The coefficients, $\mathcal{J}_{y}(\zh)$ and $\mathcal{J}_{\mu}(\zh)$, which approximate the transition between the $\mu$- and $y$-distortion regime, can be determined using a least square fit to the Green's function. We find
\beal
\label{eq:J_y}
\nonumber
\mathcal{J}_{y}(\zh)&\approx \left(1+\left[\frac{1+z}{\pot{6.0}{4}}\right]^{2.58}\right)^{-1}
\\
\mathcal{J}_{\mu}(\zh)&\approx 1-\exp\left(-\left[\frac{1+z}{\pot{5.8}{4}}\right]^{1.88}\right).
\end{align}
%
This approximation for $G_{\rm th}(\nu, z_{\rm h}, 0)$ works at $\simeq 10\%-30\%$ precision for the standard cosmology, with the residuals containing additional information about the time-dependence of the energy release at $10^4 \lesssim \zh\lesssim \pot{3}{5}$. In summary, this means that a pretty good approximation for the Green's function at all redshifts is given by 
\beal
\label{eq:Greens_all}
G^\ast_{\rm th}(\nu, z_{\rm h}, 0)&= 1.401\mathcal{J}_{\mu}(\zh)\, \mathcal{J}(\zh)\,M(\nu)
\nonumber\\
&\qquad+\frac{\mathcal{J}_{y}(\zh)}{4}\,Y_{\rm SZ}(\nu)+\frac{1-\mathcal{J}(\zh)}{4}\,G(\nu).
\end{align}
The redshift-dependence of the functions, $\mathcal{J}_{y}(\zh)$, $\mathcal{J}_{\mu}(\zh)$ and $\mathcal{J}(\zh)$ is illustrated in Fig.~\ref{fig:examples}. At $z\simeq \pot{5.3}{4}$ we find $\mathcal{J}_{y}(\zh)\simeq \mathcal{J}_{\mu}(\zh)$, justifying why $z\simeq \pot{5}{4}$ was previously used to mark the transition between $\mu$- and $y$-era \citep{Hu1993}.

It is also important to mention that the energy integral $\int [G^\ast_{\rm th}(\nu, z_{\rm h}, 0)/\rho_\gamma]\id \nu=\left[\mathcal{J}_{\mu}(\zh)\,\mathcal{J}(\zh)+\mathcal{J}_{y}(\zh)+1-\mathcal{J}(\zh)\right]\lesssim 1$. The `missing' energy can be found in the residual distortion, which is not represented well by the approximation, although it never exceeds $\simeq 16\%$ (maximum at $z\simeq \pot{7.7}{4}$). For simple estimates, Eq.~\eqref{eq:Greens_all} should thus be very useful, while for high precision the exact form of the Green's function must be used.

\subsection{Spectral distortion for some examples}
\label{sec:examples_diss}
We implemented the Green's function method as part of {\sc CosmoTherm}\footnote{{\sc CosmoTherm} is available at \url{www.Chluba.de/CosmoTherm}.}. With Eq.~\eqref{eq:final_dist} this allows very fast computation (a fraction of a second versus $\simeq10$ min -- 1 h per model) of the resulting spectral distortion for given thermal history.
In Fig.~\ref{fig:examples} we show the heating rate and resulting distortion for some illustrative cases.  
We confirmed the precision of the Green's function approach by explicitly computing the distortion for different scenarios solving the evolution equations with {\sc CosmoTherm}, finding excellent agreement [the numerical precision is much better than $\simeq 1\%$; at this level uncertainties in the background cosmology are more relevant]. 
The two annihilation scenarios assume different cross sections, representing s-wave $[\left<\sigma \rm v\right>\simeq {\rm const}]$, and p-wave annihilation $[\left<\sigma \rm v\right>\simeq T/m\propto (1+z)]$ \citep[see][for related discussion]{McDonald2001}. This p-wave scenario corresponds to a Majorana particle which either is still relativistic after freeze out [e.g., a sterile neutrino with low abundance \citep{Ho2013}], or shows $1/\rm v$ Sommerfeld-enhanced annihilation cross section \citep[e.g., see][]{Chen2013}. For a non-relativistic Majorana particle the cross section scales even faster with redshift, $\left<\sigma \rm v\right>\simeq (1+z)^2$, causing practically no energy release at late times.

Strong constraints for s-wave annihilation of dark matter can be derived from measurements of the CMB temperature anisotropies \citep{Galli2009, Huetsi2009, Huetsi2011, Planck2013params}, however, CMB anisotropy limits on the p-wave scenario are much weaker because most of the energy is liberated prior to the recombination epoch ($z\simeq 10^3$). We can see that for p-wave annihilation most of the energy is released during the $\mu$-era, while for s-wave annihilation heating at low redshifts also contributes significantly. Spectral distortions thus might allow shedding light on these cases, although degeneracies with other scenarios are expected. 

For illustration we also chose a decaying particle scenario with the main heating occurring at $z\simeq \pot{5}{4}$. This could be related to dark matter in an excited state \citep[e.g.,][]{Finkbeiner2007, Pospelov2007}, or some other, dynamically unimportant relic particle \citep[see][for more references]{Feng2010, Pospelov2010}. As explained above, in this case the distortion is not just given by a simple superposition of (pure) $\mu$ and $y$-distortion, so that this scenarios might be distinguishable from, for example, the signal created by the dissipation of acoustic modes \citep{Chluba2012, Chluba2012inflaton}, at least when the simplest shape for the primordial power spectrum is assumed. 
A more detailed forecast will be required\footnote{See \citet{Khatri2013forecast} for some recent discussion.}, and we plan  to apply our method, assuming experimental settings similar to PIXIE \citep{Kogut2011PIXIE}. 

\subsection{Cosmology-dependence of the Green's function}
The cosmology-dependence of the Green's function is mainly driven by (i) the CMB monopole temperature; (ii) the number density of baryons and electrons; (iii) the fraction of helium atoms (through the Bremsstrahlung Gaunt-factors and the effect on the number of electrons per baryon); and (iv) the expansion rate of the Universe (i.e., $T_0$, $N_{\rm eff}$ and at later stages $\Omega_{\rm m}$). Since the cosmological parameters are known with percent-level precision, the Green's function is accurate to a similar level. For detailed parameter forecasts the cosmology-dependence of the Green's function could thus lead to significant degeneracies, if differences at the level of a few percent matter. This can be easily taken into account (part of the dependencies can be taken into account by changing variables), however, since constraints on different scenarios to better than $\simeq 1\%$ precision will be challenging to obtain, we leave an in depth discussion to future work.

\section{Conclusions}
\label{sec:conclusions}
We computed the Green's function for the cosmological thermalization problem assuming standard cosmological parameters. With our method it is possible to predict the resulting spectral distortion signal for different energy release scenarios with high precision at practically no computational cost (see Fig~\ref{fig:examples} for some examples). This approach is now part of {\sc CosmoTherm} and will be very useful for detailed forecasts of possible future constraints on the pre-recombination thermal history of our Universe derived from measurements of the CMB spectrum. 

For our computations we assumed that the energy injection process itself does not lead to significant injection of photons. Generalization is straightforward, however, the associated Green's function is not only a function of injection redshift but also injection frequency, which is computationally more challenging. Since commonly discussed energy release scenarios do not fall into this category, we leave a more in depth discussion to future work.

For illustration of the method we discussed some simple energy release scenarios (see Sect.~\ref{sec:examples_diss}). In the future, it will be very important to ask how degenerate different scenarios are. For example, energy release related to dark matter annihilation with s-wave cross section is expected to give rise to a distortion that is practically degenerate with those caused by dissipation of adiabatic modes in the early Universe, if the spectral index $\nS$ of the primordial power spectrum at wavenumber $k\gtrsim 1\,\Mpc^{-1}$ is scale-invariant \citep[e.g. see][]{Chluba2011therm}. One the other hand, recent constraints from the Planck satellite \citep{Planck2013params} show that at large scales $\nS\simeq 0.96$. Extrapolating this to small scales thus suggests that the shape of the distortion ought be different in the two cases.
Our computations also show that thermal histories with decaying relics might be constrainable if the lifetime of the particle is $t_{\rm X} \simeq \pot{2.6}{8}\,{\rm sec}-\pot{2.2}{11}\,{\rm sec}$. 
We look forward to exploring these possibilities in the future.

\small

\section*{Acknowledgments}
JC thanks Yacine Ali-Ha\"{i}moud, Rishi Khatri and Rashid Sunyaev for useful discussions and comments on the paper. He also thanks Rajat Thomas for great hospitality during his visit to Amsterdam, where most of this work was finished.
Use of the GPC supercomputer at the SciNet HPC Consortium is acknowledged. SciNet is funded by: the Canada Foundation for Innovation under the auspices of Compute Canada; the Government of Ontario; Ontario Research Fund - Research Excellence; and the University of Toronto. JC was support by the grants DoE SC-0008108 and NASA NNX12AE86G.

%
%

\bibliographystyle{mn2e}
\bibliography{Lit}

\end{document}